\documentclass[letterpaper, 10 pt, conference]{IEEEtran}
\IEEEoverridecommandlockouts
\usepackage{cite}
\usepackage{amsmath,amssymb,amsfonts}
\usepackage{algorithmic}
\usepackage{algorithm}
\usepackage{array}
\usepackage{graphicx}
\usepackage{textcomp}
\usepackage{xcolor}
\usepackage{geometry}
\usepackage{comment}
\usepackage{stfloats}
\usepackage{url}
\usepackage{verbatim}
\usepackage{enumerate}

\usepackage{enumitem}
\usepackage{multirow}
\usepackage{diagbox}
\usepackage{mathtools, cuted}
\usepackage{lipsum, color}
\usepackage{standalone}

\usepackage{subcaption}

\usepackage{tikz}
\usepackage{pgfplots}

\begin{document}
	\newgeometry{left=19.1mm,right=19.1mm,top=25.4mm,bottom=19.1mm}	
	\title{\LARGE \bf Experimental Implementation and Validation of Predictor-Based CACC for Vehicular Platoons With Distinct Actuation Delays}
	
	\author{Amirhossein Samii$^1$, Redmer de Haan$^2$, and Nikolaos Bekiaris-Liberis$^1$
		\thanks{			
			$^1$A. Samii and N. Bekiaris-Liberis are with the Department of Electrical \& Computer Engineering, Technical University of Crete, Chania, 73100, Greece. Email
			addresses: 
			{\tt\small asamii@tuc.gr} and {\tt\small bekiaris-liberis@ece.tuc.gr.}
			
			$^2$R. de Haan is with the Department of Mechanical Engineering, Eindhoven University of Technology, Eindhoven, 5600 MB Eindhoven, Netherlands. Email address: 
			{\tt\small r.d.haan@tue.nl.}}%
	}
	
	\maketitle
	
	\begin{abstract}
		We provide experimental validation, in a pair of vehicles, of a recently introduced predictor-based cooperative adaptive cruise control (CACC) design, developed for achieving delay compensation in heterogeneous vehicular platoons subject to long actuation delays that may be distinct for each individual vehicle. We provide the explicit formulae of the control design that is implemented, accounting for the effect of zero-order hold and sampled measurements; as well as we obtain vehicle and string stability conditions numerically, via derivation of the transfer functions relating the speeds of pairs of consecutive vehicles. We also present consistent simulation results for a platoon with a larger number of vehicles, under digital implementation of the controller. Both the simulation and experimental results confirm the effectiveness of the predictor-based CACC design in guaranteeing individual vehicle stability, string stability, and tracking, despite long/distinct actuation delays. 
	\end{abstract}
	\section{Introduction}
	\subsection{Motivation}
	Actuation delays negatively affect performance of adaptive cruise control (ACC)- and CACC-equipped vehicles, see, for example, \cite{predictor_feedback1,speed_error2,b7,prediction1,italian,L_stability,dy1,b38}. Such delays may be, typically, different for each individual vehicle \cite{b31,speed_error2,predictor_based,speed_error3} and they can be long \cite{predictor_feedback2,dutch,dutch1,b14,predictor_feedback3,b13,b29,dy2,b38,speed_error3}. For this reason there are predictor-based ACC/CACC design methodologies that aim at long/distinct actuation delays compensation \cite{nominal_control_1,predictor_feedback1},\cite{predictor_feedback2,b31},\cite{b6,speed_error2,dutch2,b14,prediction1,predictor_feedback3,predictor_based,b13,dy1,b38,b50},\cite{speed_error3,distinc_input_delay2}. Such methods are typically validated in simulation, including testings with real data. However, to the best of our knowledge only \cite{dutch2} and \cite{dy1} experimentally validate ACC/CACC designs with delay compensation capabilities. Motivated by this, in the present work we complement \cite{dutch2}, \cite{dy1} in experimentally implementing and validating the predictor-based CACC design developed in \cite{predictor_based}, which, differently from \cite{dutch2}, \cite{dy1}, can be applied to vehicles with distinct/long actuation delays, while guaranteeing provably vehicle stability and string stability, and providing a systematic/modular approach to design of delay-compensating ACC/CACC laws, as it can be combined with any given nominal, delay-free ACC/CACC design. Thus, such experimental implementation and validation paves the way for potential large-scale implementation and deployment of delay-compensating, predictor-based ACC/CACC designs, with potential benefits in terms of fuel consumption, safety, and comfort of vehicular platoons. 
	\subsection{Literature}
	Among the predictor-based ACC/CACC designs for vehicular platoons with long actuation delays \cite{nominal_control_1,predictor_feedback1},\cite{predictor_feedback2,b31,b6},\cite{speed_error2,dutch2,b14,prediction1,predictor_feedback3,predictor_based,b13,dy1,b38,b50},\cite{speed_error3,distinc_input_delay2}, distinct and long actuation delays are considered in \cite{b31,speed_error3,distinc_input_delay2}. However, as compared with these works, the approach from \cite{predictor_based}, which we experimentally implement and validate here, is applicable to heterogeneous vehicles with long and distinct actuation delays, while providing a constructive/systematic approach for establishing both vehicle and string stability. Our work is also relevant to experimental implementations of predictor-based controllers, which have been performed in \cite{predictor-based_implementation1} for low level engine control, in \cite{predictor-based_implementation2} for control of a robot manipulator, and in \cite{predictor-based_implementation3} for control of an unmanned aerial vehicle. As described above, only in \cite{dutch2} and \cite{dy1} experimental implementations and validations of predictor-based ACC/CACC designs are considered. Our work complements \cite{dutch2} and \cite{dy1} in that we experimentally implement and validate a new predictor-based CACC design that achieves complete delay compensation and which is accompanied with provable guarantees of vehicle/string stability and tracking, for vehicular platoons in which each vehicle may feature a different actuation delay, which is typically the case in practice. 
	\subsection{Contributions}
	In the present paper we provide the explicit formulae of the predictor-based CACC design from \cite{predictor_based} when it is applied through zero-order hold employing sampled measurements. For such a sampled-data implementation we obtain vehicle and string stability conditions numerically, via derivation of the respective transfer functions relating the speeds of pairs of consecutive vehicles. We present consistent simulation results for a platoon of five vehicles, under digital implementation of the controller. We then present the description of the experimental setup, including the measurements/sensors employed, the parameters of control implementation, and the parameters identified for the model of the vehicles; as well as we present the results of the experimental implementation and validation of our controller in a pair of vehicles. Although the vehicles employed are small two-seated electric vehicles, the experiments involve all practical phenomena that would appear in control of larger vehicles, while, in particular, longer actuation delays can be artificially introduced \cite{dutch3}. The experimental results confirm that vehicle stability and\newgeometry{left=19.1mm,right=19.1mm,top=19.1mm,bottom=19.1mm} \noindent string stability in speed error propagation are achieved despite long and distinct actuation delays affecting the vehicles, even in the hardest case (with respect to computation of the predictor states) in which the ego vehicle has a larger actuation delay (as compared with the preceding vehicle). The simulation and experimental results are consistent with the theory and illustrate the potential significance of actual implementation of predictor-based CACC designs. 
	\subsection{Organization}
	In Section II we present the model considered for control design, together with the basic, continuous predictor-based CACC law; as well as we provide the formulae of our design when it is applied with zero-order hold employing sampled measurements. In Section III we derive numerical conditions for vehicle and string stability, under digital implementation of the controller, as well as we present simulation results. In Section IV we present experimental results. We provide concluding remarks in Section V.
	
	\section{CACC for Heterogeneous Platoons With Distinct Actuation Delays}
	\subsection{Vehicle Model and Available Measurements}\
	\textit{a) Vehicle dynamics:} We consider a heterogeneous string of vehicles (see Fig. 1) each one modeled by the following third-order, linear system with distinct actuator delays that describes vehicle dynamics (see, e.g.,\cite{predictor_feedback3},\cite {dy1},\cite{dy2})
	\begin{align}
		\dot{s}_i(t) =&\nobreakspace v_{i-1}(t) - v_i(t),\label{dy1}\\	
		\dot{v}_i(t) =&\nobreakspace a_i (t),\label{dy2}\\
		\dot{a}_i(t) =&\nobreakspace -\frac{1} {\tau_i} a_i (t)+\frac{1} {\tau_i}u_i(t-D_i),\label{dy3}
	\end{align}
	$i = 1,...,N$, where $s_i=x_{i-1} - x_i - l$ and $x_i$ is the position of vehicle $i$ and $l$ is its length, $v_i$ is vehicle speed, $a_i$ is vehicle acceleration, $\tau_i$ is lag, capturing, engine dynamics, $u_i$ is the individual vehicle’s control variable, $D_i \geq 0$ are input delays, and $t \geq 0$ is time. Note that for the leading vehicle we assume similarly that it has the same type of third-order dynamics as the rest of the vehicles\footnote{The design can be modified in a straightforward manner when this is not true.}. We adopt the convention that $v_0 = v_l$ and $a_0 = a_l$ are the speed and acceleration of the string leader, respectively.
	\begin{figure}[!h]
		\begin{center}
			\includegraphics[width = 8cm, height = 2cm]{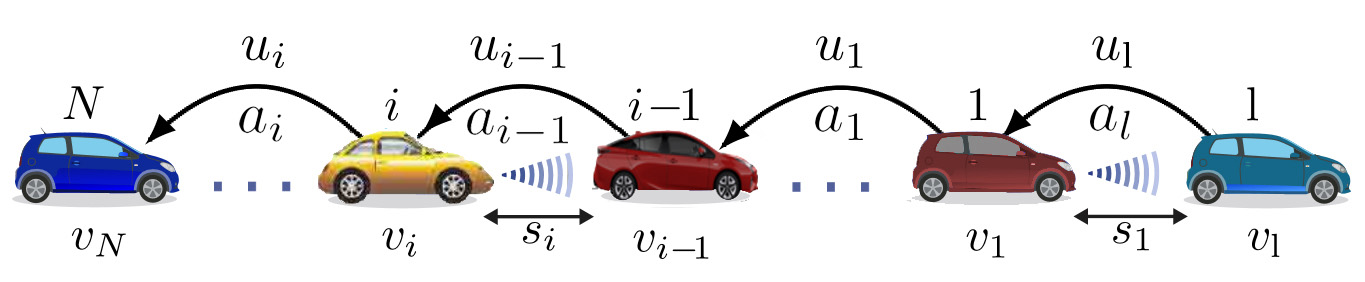}
			\caption{Platoon of $N+1$ heterogeneous vehicles following each other in a single lane without overtaking. The dynamics of each vehicle $i = 1,..., N$ are governed by system (\ref{dy1})--(\ref{dy3}). Each vehicle can measure its own speed, the relative speed with the preceding vehicle, and the spacing with respect to the preceding vehicle. The control input and acceleration of each vehicle is communicated to the following vehicle via V2V communication.}
		\end{center}
		\label{Fig1}
	\end{figure}
	
	\textit{b) Available measurements:} For the platoons considered here the measurements available to the ego vehicle $i$ are its own spacing $s_i$, speed $v_i$, acceleration $a_i$, and control input $u_i$, as well as the speed of the preceding vehicle $v_{i-1}$. This information is obtained through on-board sensors. Furthermore, the control input of the preceding vehicle and its acceleration are also available, and they are denoted by $u_{i-1}$ and $a_{i-1}$, respectively. These measurements are transmitted from the preceding vehicle, through vehicle-to-vehicle (V2V) communication. Note that the speed of the preceding vehicle, $v_{i-1}$, can also be obtained via V2V communication.
	
	\subsection{Continuous Predictor-Based CACC Design}
	\textit{a) Nominal control design:} Without input delay, the following control strategy is constructed (see, e.g.,\cite {nominal_control_1,nominal_control_2})
	\begin{equation}
		\begin{aligned}
			u_i(t)= &\nobreakspace \tau_i\alpha_i\left( \frac{s_i(t)-r_i} {h_i}-v_i(t)\right)+\tau_ib_i(v_{i-1}(t)-v_i(t))\\ 
			&+\tau_ic_i\left( a_{i-1}(t)-a_i(t)\right) ,
			\label{CLN}
		\end{aligned}
	\end{equation}
	where $\alpha_i>0$, $b_i>0$, and $c_i\geq0$ are design parameters, $h_i > 0$ is time-headway, and $r_i>0$ is inter-vehicle distance at standstill.
	
	\textit{b) Predictor-based CACC:} The predictor-based control laws for system (\ref{dy1})--(\ref{dy3}) in continuous time are given by (see \cite{predictor_based})
	\begin{align}
		u_i(t)= &\nobreakspace \frac{\tau_i\alpha_i}{h_i} q_{i,1}(t)-\tau_i(\alpha_i+b_i)q_{i,2}(t)+\tau_i b_i q_{i,3}(t)\nonumber\\
		&+\tau_ic_i\left(q_{i,5}(t)-q_{i,4}(t)\right) ,\label{CL}
	\end{align}
	where
	\begin{align}
		q_{i}(t)= &\nobreakspace {\rm e}^{\Gamma_i D_i} \bar{x}_i(t)+ \int_{t-D_i}^{t} {\rm e}^{\Gamma_i(t-\theta)}B_iu_{i}(\theta)\, d\theta \nonumber\\
		&+\int_{t-D_{i-1}}^{t} {\rm e}^{\Gamma_i(t+D_i-\theta-D_{i-1})}B_{1i}u_{i-1}(\theta)\, d\theta, \label{PF}
	\end{align}
	with
	\begin{align}
		q_{i}=& \begin{bmatrix} 
			q_{i,1}  \\
			q_{i,2} \\
			q_{i,3} \\
			q_{i,4} \\
			q_{i,5} \\
		\end{bmatrix},\quad
		\bar{x}_{i}= \begin{bmatrix} 
			s_{i}-r_i  \\
			v_{i}  \\
			v_{i-1} \\
			a_{i} \\
			a_{i-1} \\
		\end{bmatrix},\label{C1}\\
		B_i=& \begin{bmatrix} 
			0&0&0&\frac{1}{\tau_{i}}&0
		\end{bmatrix}^{\rm T},
		B_{1i}= \begin{bmatrix} 
			0&0&0&0&\frac{1}{\tau_{i-1}}
		\end{bmatrix}^{\rm T},\label{C2}\\
		\Gamma_i=& \begin{bmatrix} 
			0 & -1 & 1 & 0 & 0 \\
			0 & 0 & 0 & 1 & 0\\
			0 & 0 & 0 & 0 & 1\\
			0 & 0 & 0 & -\frac{1}{\tau_{i}} & 0\\
			0 & 0 & 0 & 0 & -\frac{1}{\tau_{i-1}}\\
		\end{bmatrix}.\label{C3}
	\end{align}
	\subsection{Predictor-Based CACC Design for Distinct Delays Implemented With Zero-Order Hold}
	Let $T_s>0$ be the sampling period, with $D_i\geq 0$ be given and $l_i \in Z^+$ such that $D_i=l_iT_s$. Define the following quantities
	\begin{align}
		u_i(t)=&u_{i_k},\quad t \in [kT_s, (k+1)T_s),\quad k\in Z^+,\\
		\bar{x}_{i_k}=&\bar{x}_i(kT_s),\\
		\bar{x}_{{i,{\rm m}}_k}=&\bar{x}_i(kT_s-D_{{\rm c},i}),
	\end{align}
	where $D_{{\rm c},i}=l_{i,{\rm c}}T_s$ is communication delay. Then, the control inputs (\ref{CL}) implemented with zero-order-hold are (see also \cite{Discrete_predictor})
	\begin{align}
		u_{i_k}=& \frac{\tau_i\alpha_i}{h_i}  q_{{i,1}_k} -\tau_i(\alpha_i+b_i)q_{{i,2}_k}+\tau_i b_i q_{{i,3}_k}\nonumber\\
		+&\tau_ic_i\left(q_{{i,5}_k}-q_{{i,4}_k}\right) ,\label{CLD}
	\end{align}
\begin{align}
		q_{i_k}=&{\rm e}^{\Gamma_iD_i}\bar{x}_{i_k}+ \sum_{j=1}^{l_i}Q_{i_j}B_iu_{i_{k-j}}\nonumber \\ 
		+&{\rm e}^{\Gamma_i(D_i-D_{i-1})}\sum_{j=1}^{l_{i-1}}Q_{{i-1}_j}B_{1i}u_{{i-1,{\rm m}}_{k-j}}\label{8}, 
	\end{align}
	where 
	\begin{align}
		Q_{{i}_j}=& {\rm e}^{\Gamma_ijT_s}\int_{0}^{T_s}{\rm e}^{-\Gamma_i \theta}d\theta,\quad j=1,...,l_i,\label{9_0}\\
		Q_{{i-1}_j}=& {\rm e}^{\Gamma_ijT_s}\int_{0}^{T_s}{\rm e}^{-\Gamma_i \theta}d\theta,\quad j=1,...,l_{i-1}\label{9},
	\end{align}
	and
	\begin{align}
		q_{i_k}=& \begin{bmatrix} 
			q_{{i,1}_k}  \\
			q_{{i,2}_k} \\
			q_{{i,3}_k} \\
			q_{{i,4}_k} \\
			q_{{i,5}_k} \\
		\end{bmatrix},\quad
		\bar{x}_{i_k}= \begin{bmatrix} 
			s_{i_k}-r_i  \\
			v_{i_k}  \\
			v_{{i-1,{\rm m}}_k} \\
			a_{{i}_k} \\
			a_{{i-1,{\rm m}}_k} \\
		\end{bmatrix},\label{C4}
	\end{align}
	\setlength\arraycolsep{2pt}
	\begin{strip}
		\begin{align}
			&{\rm e}^{\Gamma_i D_i}=\begin{bmatrix} 
				1 & -D_i & D_i & \tau_i^2-D_i\tau_i-\tau_i^2{\rm e}^{\frac{-D_i}{\tau_i}} & -\tau_{i-1}^2+D_i\tau_{i-1}-\tau_{i-1}^2{\rm e}^{\frac{-D_i}{\tau_{i-1}}} \\
				0 & 1 & 0 & \tau_i-\tau_i{\rm e}^{\frac{-D_i}{\tau_i}} & 0\\
				0 & 0 & 1 & 0 & \tau_{i-1}-\tau_{i-1}{\rm e}^{\frac{-D_i}{\tau_{i-1}}}\\
				0 & 0 & 0 & {\rm e}^{\frac{-D_i}{\tau_i}} & 0\\
				0 & 0 & 0 & 0 & {\rm e}^{\frac{-D_i}{\tau_{i-1}}}\\
			\end{bmatrix},\\
			&{\rm e}^{\Gamma_i \left( D_i-D_{i-1}\right) }=\nonumber \\
			&\begin{bmatrix} 
				1 & -\left( D_i-D_{i-1}\right) & D_i-D_{i-1} & \tau_i^2-\left( D_i-D_{i-1}\right)\tau_i-\tau_i^2{\rm e}^{\frac{-\left( D_i-D_{i-1}\right)}{\tau_i}} & -\tau_{i-1}^2+\left( D_i-D_{i-1}\right)\tau_{i-1}-\tau_{i-1}^2{\rm e}^{\frac{-\left( D_i-D_{i-1}\right)}{\tau_{i-1}}} \\
				0 & 1 & 0 & \tau_i-\tau_i{\rm e}^{\frac{-\left( D_i-D_{i-1}\right)}{\tau_i}} & 0\\
				0 & 0 & 1 & 0 & \tau_{i-1}-\tau_{i-1}{\rm e}^{\frac{-\left( D_i-D_{i-1}\right)}{\tau_{i-1}}}\\
				0 & 0 & 0 & {\rm e}^{\frac{-\left( D_i-D_{i-1}\right)}{\tau_i}} & 0\\
				0 & 0 & 0 & 0 & {\rm e}^{\frac{-\left( D_i-D_{i-1}\right)}{\tau_{i-1}}}\\
			\end{bmatrix},
		\end{align}
		\begin{align}
			\int_{0}^{T_s}{\rm e}^{-\Gamma_i \theta}d\theta=&\begin{bmatrix} 
				T_s & \frac{T_s^2}{2} & -\frac{T_s^2}{2} & \frac{\tau_i}{2}\left(T_s^2+2T_s\tau_i-2\left(-1+{\rm e}^{\frac{T_s}{\tau_i}} \right)\tau_i^2  \right)  & \frac{\tau_{i-1}}{2}\left(T_s^2+2T_s\tau_{i-1}-2\left(-1+{\rm e}^{\frac{T_s}{\tau_{i-1}}} \right)\tau_{i-1}^2  \right) \\
				0 & T_s & 0 & \tau_i\left( T_s+\tau_i-\tau_i{\rm e}^{\frac{T_s}{\tau_i}}\right)  & 0\\
				0 & 0 & T_s & 0 & \tau_{i-1}\left( T_s+\tau_{i-1}-\tau_{i-1}{\rm e}^{\frac{T_s}{\tau_{i-1}}}\right)\\
				0 & 0 & 0 & \tau_i\left(-1+{\rm e}{\frac{T_s}{\tau_i}} \right)  & 0\\
				0 & 0 & 0 & 0 & \tau_{i-1}\left(-1+{\rm e}{\frac{T_s}{\tau_{i-1}}} \right)\\
			\end{bmatrix}.\label{Co_3}
		\end{align}
	\end{strip}
\vspace*{-\baselineskip}
	\section{Numerical Investigation of String Stability in Discrete Time And Simulation Results}
	We start providing the definition of string stability employed. A platoon of vehicles indexed by $i=1,...,N,$ following each other within one lane without overtaking, is $\mathcal{L}_2$ string stable with reference to speed errors if the following condition holds (see, e.g., \cite{b17})
	\begin{align}
		\mathop{\sup}_{\omega}|G_i\left( {\rm e}^{j\omega T_s}\right) |\leq 1, \quad\quad i=1,...,N,\label{G}
	\end{align}
	where $G_i\left( {\rm e}^{j\omega T_s}\right)$ denotes the transfer function between the $i$-th vehicle's speed and the speed of its preceding vehicle $i-1$. Here we study $\mathcal{L}_2$ string stability with respect to speed errors propagation, as this is the most commonly used definition, see, for example, \cite {speed_error1,dutch,speed_error2,speed_error3}. 
	The derivation of $G_i$ can be found in Appendix A. Based on (\ref{Gz}) we perform numerical studies to investigate the range of control parameters for which vehicle and string stability hold, also depending on the sampling period $T_s$.
	
	
	\subsection{Choice of Control Parameters}
	We numerically analyze the $\mathcal{L}_2$ string stability properties of the closed-loop system concerning the propagation of speed errors. The transfer function $G_i(z)$ we consider, shown in (\ref{Gz}), corresponds to the closed-loop transfer function obtained in \cite{predictor_based} and displayed in (\ref{A.24}) (for the reader's convenience) after applying the Tustin approximation \cite{tustin}. This is only an approximation of the actual transfer function one obtains in a closed-loop system consisting of (\ref{dy1})--(\ref{dy3}), subject to zero-order hold, under (\ref{CLD})--(\ref{Co_3}). We use it however for simplicity as derivation of the actual transfer function would result in a quite complicated expression, which would make it quite difficult to obtain string stability conditions even numerically.  The top plot of Fig.~2 depicts $\sup_{\omega}|G_1({\rm e}^{j\omega T_s})|$ as a function of $\alpha_1$ and $b_1$ for choices of parameters corresponding to vehicles 0, 1 from Table~\ref{table2}. The region between the red lines indicates where $\mathcal{L}_2$ string stability holds. In addition, the bottom plot of Fig.~2 demonstrates $\sup_{\omega}|G_1({\rm e}^{j\omega T_s})|$ as a function of $D_1-D_{0}$. 
	
	We note that for vehicle stability, under the zero-order hold implementation (\ref{CLD})--(\ref{Co_3}), we can use the condition in Corollary 3.4 from [17], which gives the (exact) condition that the eigenvalues of matrix 
	\begin{align}
		{\rm e}^{\bar{A}_i T_s}\left(I+\int_0^{T_s}{\rm e}^{-\bar{A}_i w}dw\bar{B}_i\bar{K}_i\right)\label{Magnitude},
	\end{align}
	are within the unit circle, where 
	\begin{align}
		\bar{A}_i=&\begin{bmatrix} 0& -1& 0\\0& 0& 1\\0& 0& -\frac{1}{\tau_i}\end{bmatrix},\quad 
		\bar{B}_i=\begin{bmatrix}0\\0\\ \frac{1}{\tau_i}\end{bmatrix}, \\
		\bar{K}_i=&\begin{bmatrix}\tau_i\frac{\alpha_i}{h_i} & -\tau_i(\alpha_i+b_i) &-\tau_ic_i\end{bmatrix}.
	\end{align}
	This holds for the model/control parameters in Table \ref{table2}. Furthermore, for the control/model parameters of vehicle~1, we show in Fig. 3 the magnitude of the maximum eigenvalue of matrix (\ref{Magnitude}) as function of $T_s$. We observe that vehicle stability is preserved for $T_s<1.67$.
	\begin{figure}[!htb]
		\begin{center}
			\includegraphics[width = 8cm, height = 6cm]{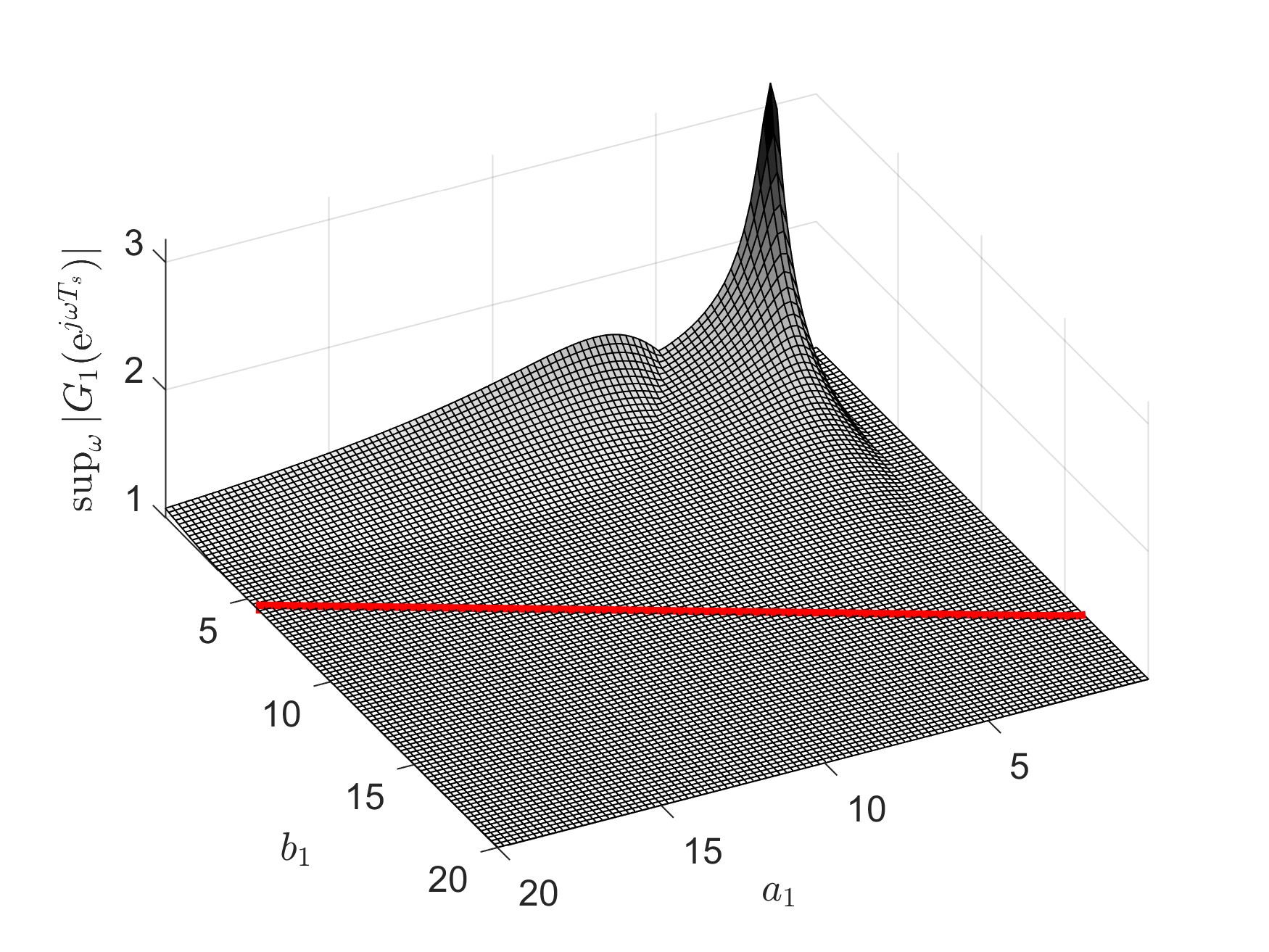}
			\includegraphics[width = 8cm, height = 6cm]{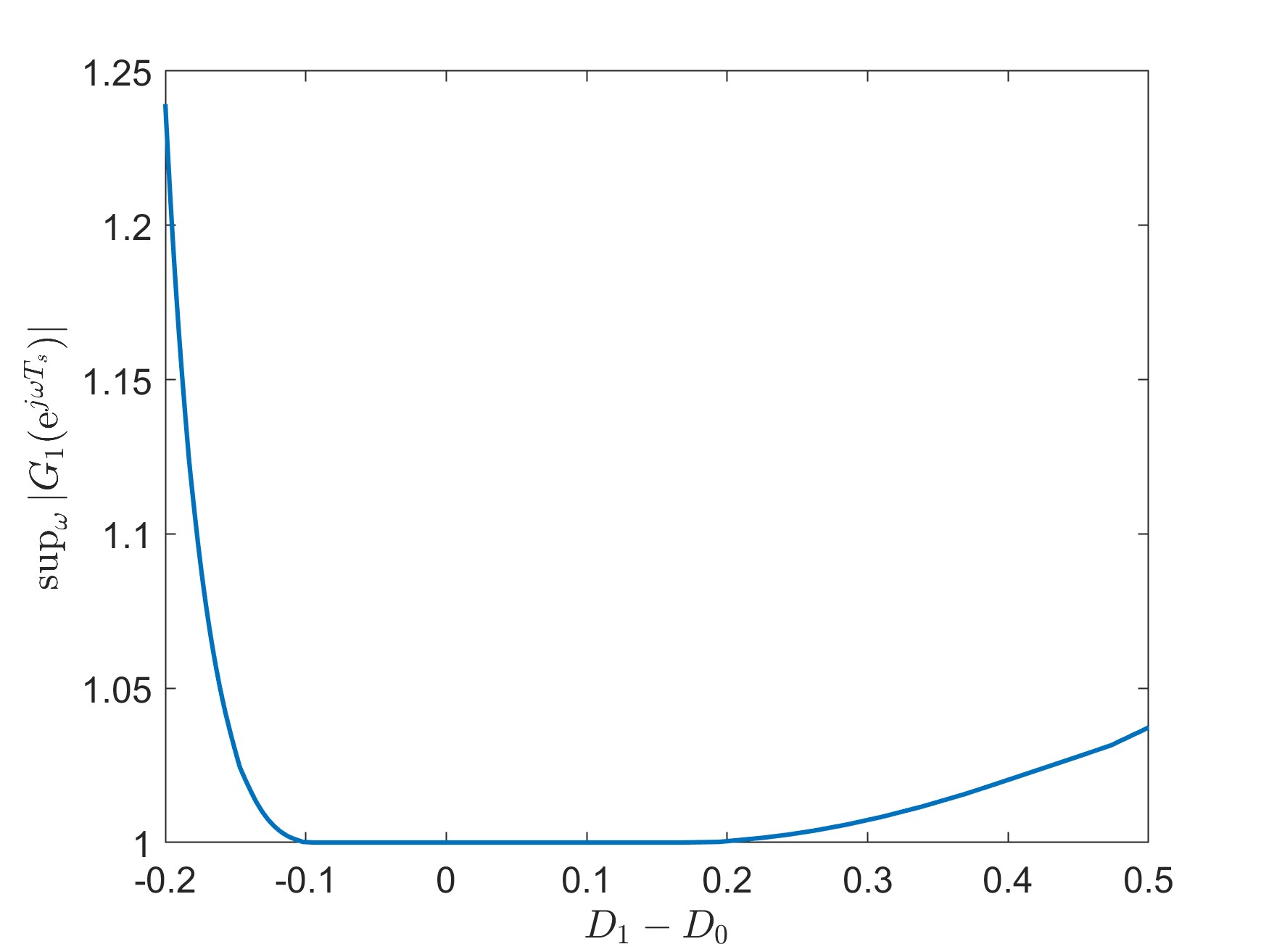}
			\caption{Magnitude $\sup_{\omega}|G_1({\rm e}^{j\omega T_s})|$ corresponding to transfer function (\ref{Gz}), as a function of $\alpha_1$ and $b_1$ for choices of parameters corresponding to vehicles 0, 1 from Table \ref{table2} (top) and $\sup_{\omega}|G_1({\rm e}^{j\omega T_s})|$ as a function of $D_1-D_{0}$ (bottom).}
		\end{center}
		\label{Fig5}
	\end{figure}
	\begin{figure}[!htb]
		\begin{center}
			\includegraphics[width = 8cm, height = 6cm]{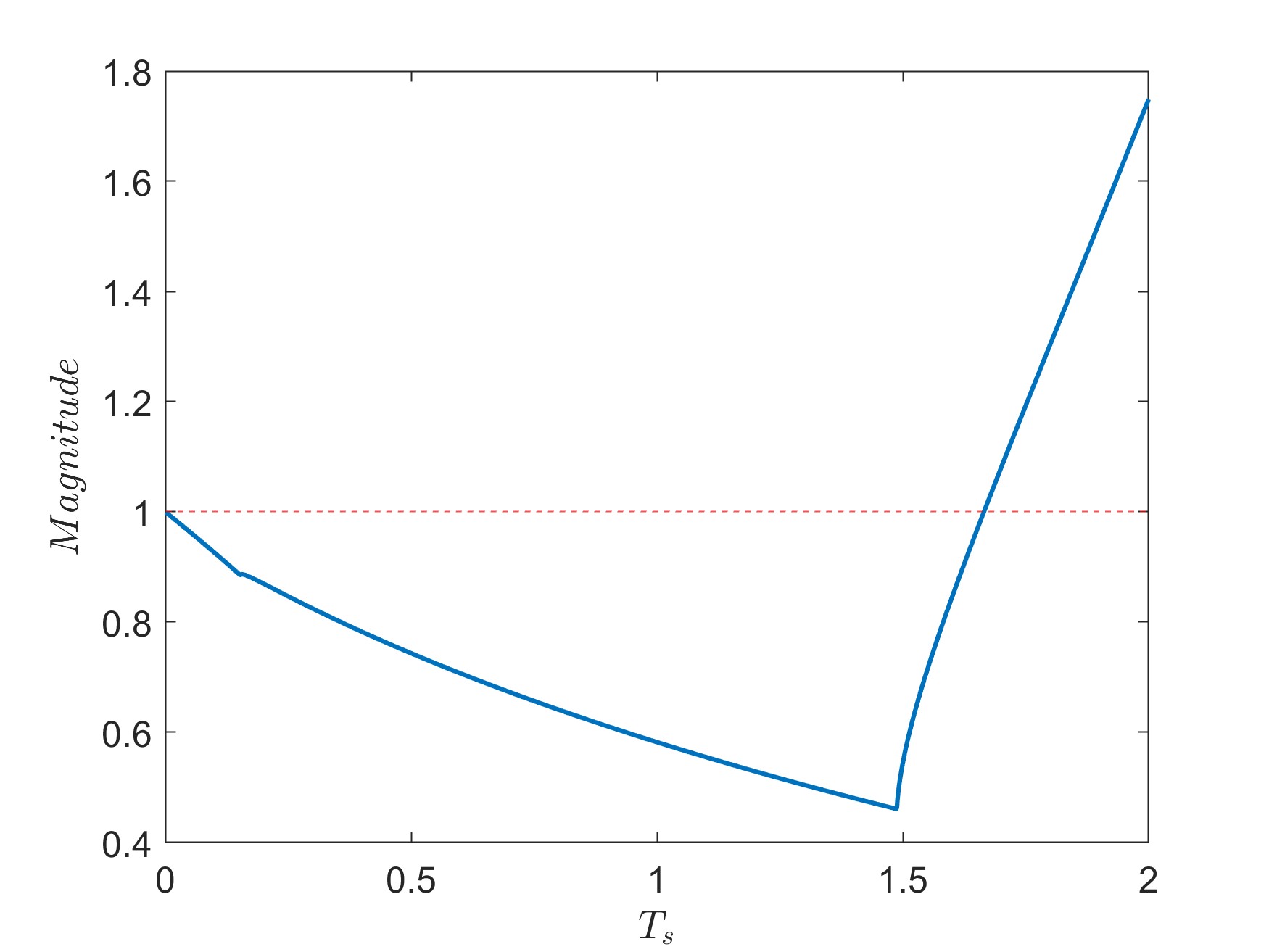}
			\caption{Magnitude of the maximum eigenvalue of matrix (\ref{Magnitude}) for $i=1$ as function of $T_s$.}
		\end{center}
		\label{Figmag}
	\end{figure}
	\subsection{Simulation Results}
	For a platoon of five vehicles with third-order dynamics given by (\ref{dy1})--(\ref{dy3}), we consider a case with parameters in Tables \ref{table2} and \ref{table3} taken from \cite{predictor_based,dutch,dutch2}. We consider a scenario in which $a_i(0)=0$; $u_{i}(s) \equiv 0$, for each vehicle $i$. In Fig. 4 we set $v_{i_0} = 10 \left(\frac{m}{s} \right)$ and $v_{{\rm l}_0} = 9 \left(\frac{m}{s} \right) $; $s_{i_0} = h_{i}v_{i_0}+r_i$~$(m)$ and $s_{1_0} = 9.5 +r_1$ $(m)$. Additionally, we add noise to all measurements of acceleration, velocity, spacing, and control input of preceding vehicle, to more closely emulate experimental implementations. We observe that the platoon is $\mathcal{L}_2$ string stable, but it may not necessarily be $\mathcal{L}_{\infty}$ string stable, which may result in the small overshoots observed (that may appear also due to noise). We note that the nominal controller (\ref{CLN}), without a delay compensation mechanism, would result in a highly oscillatory, string (and even vehicle) unstable behavior as actuation delays increase.
	
	\begin{table}[!ht]
		\centering
		\caption{The parameters employed for the numerical/experimental investigations presented in Sections III and IV.}
		\scalebox{0.9}{\begin{tabular}{||c| c c c c c||} 
				\hline
				\backslashbox{Parameters}{Vehicle No.} & 0 & 1 & 2 & 3 & 4\\ 
				\hline\hline
				$\tau_{i}$ & $0.067$ s & $0.067$ s & $0.1$ s & $0.2$ &0.15\\ 
				\hline
				$h_i$& $-$ & $1$ s & $1$ s & $0.8$ s & $0.8$ s\\ 
				\hline
				$D_i$&$0.15$ s & $0.3$ s & $0.6$ s & $0.4$ s & $0.3$ s\\
				\hline
				$D_{{\rm c},i-1}$&$-$ & $0.02$ s & $0.05$ s & $0.04$ s & $0.03$ s\\
				\hline
				$\alpha_i$&$-$  & $7.5$  & $7.5$ & $5$  &$5$\\
				\hline
				$b_i$&$-$  & $12.5$  & $12.5$  &  $10$ &$10$\\
				\hline
				$c_i$&$-$  & $0$  & $0$  & $1$  &$1$\\
				\hline
		\end{tabular}}
		\label{table2}
	\end{table}
	
	\begin{table}[!ht]
		\centering
		\caption{The parameters employed for numerical/experimental illustration.}
		\begin{tabular}{||c c ||} 
			\hline
			Parameter & Value \\ 
			\hline\hline
			$T_{s}$ & $0.01$ s \\
			\hline
			Simulation step & $0.001$ s \\
			\hline
			$r_i$ & $10$ m  \\
			\hline
		\end{tabular}
		\label{table3}
	\end{table}
	\begin{figure}[!htb]
		\begin{center}
			\includegraphics[width = 8cm, height = 6cm]{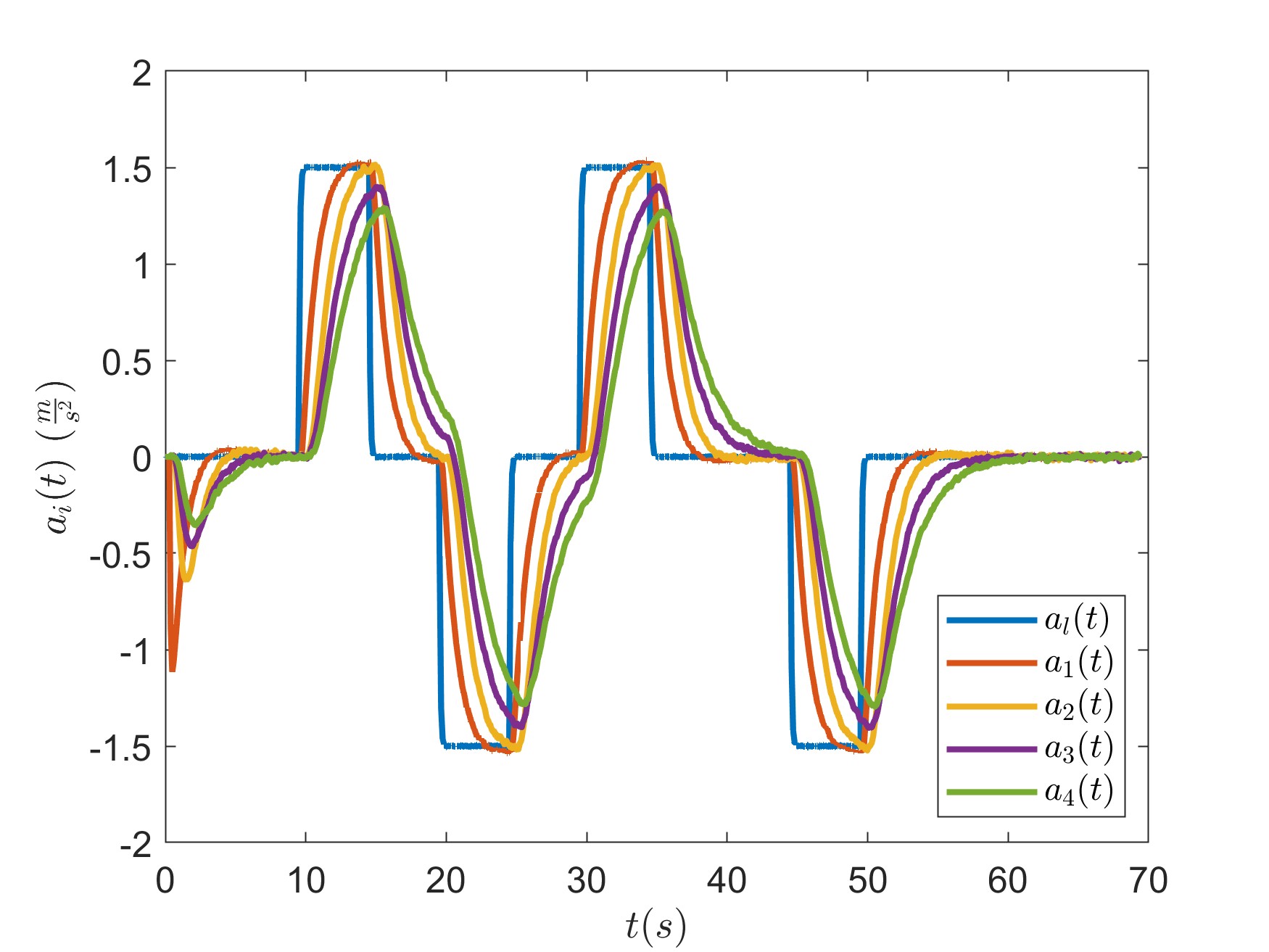}
			\includegraphics[width = 8cm, height = 6cm]{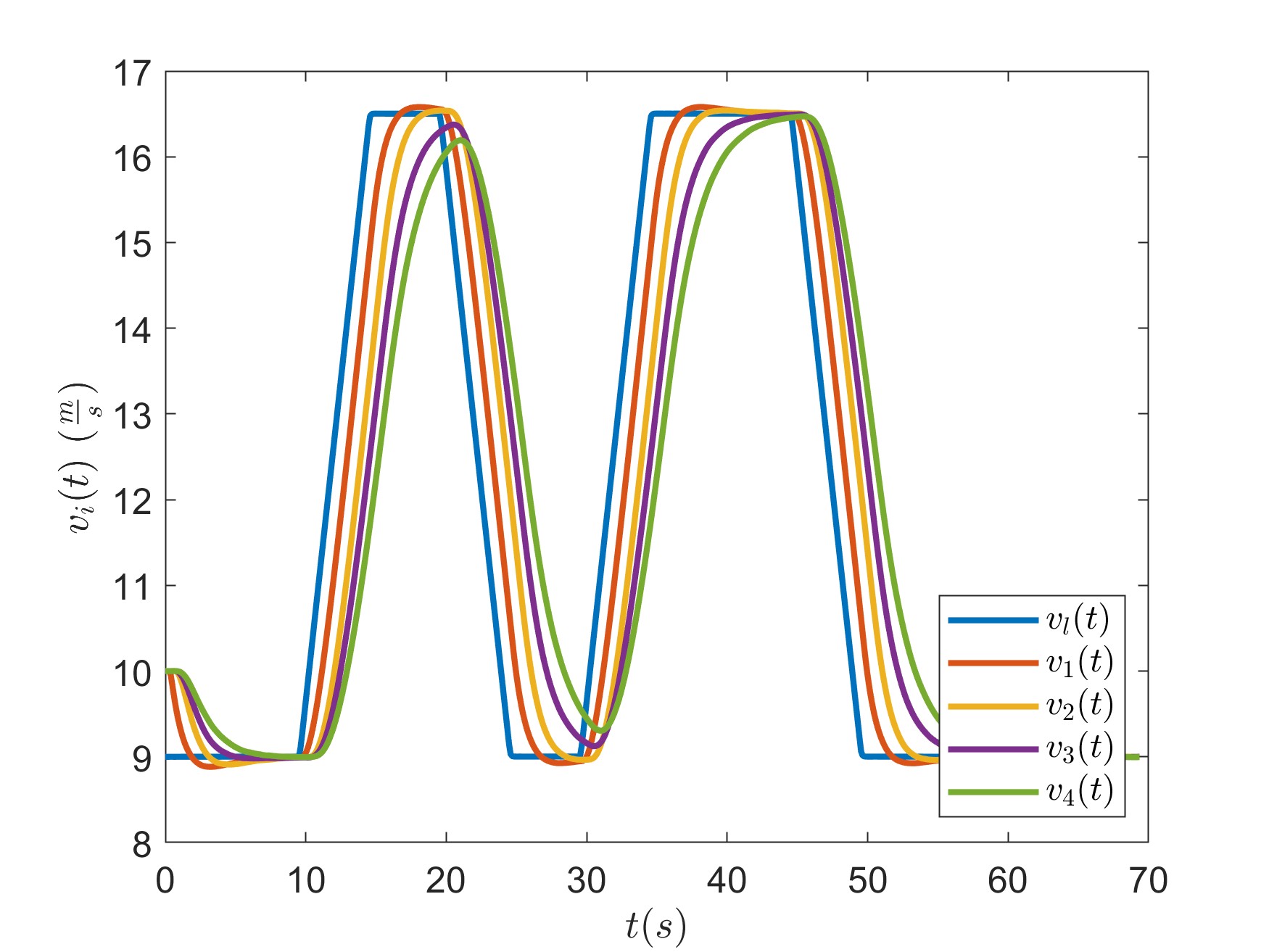}
			\includegraphics[width = 8cm, height = 6cm]{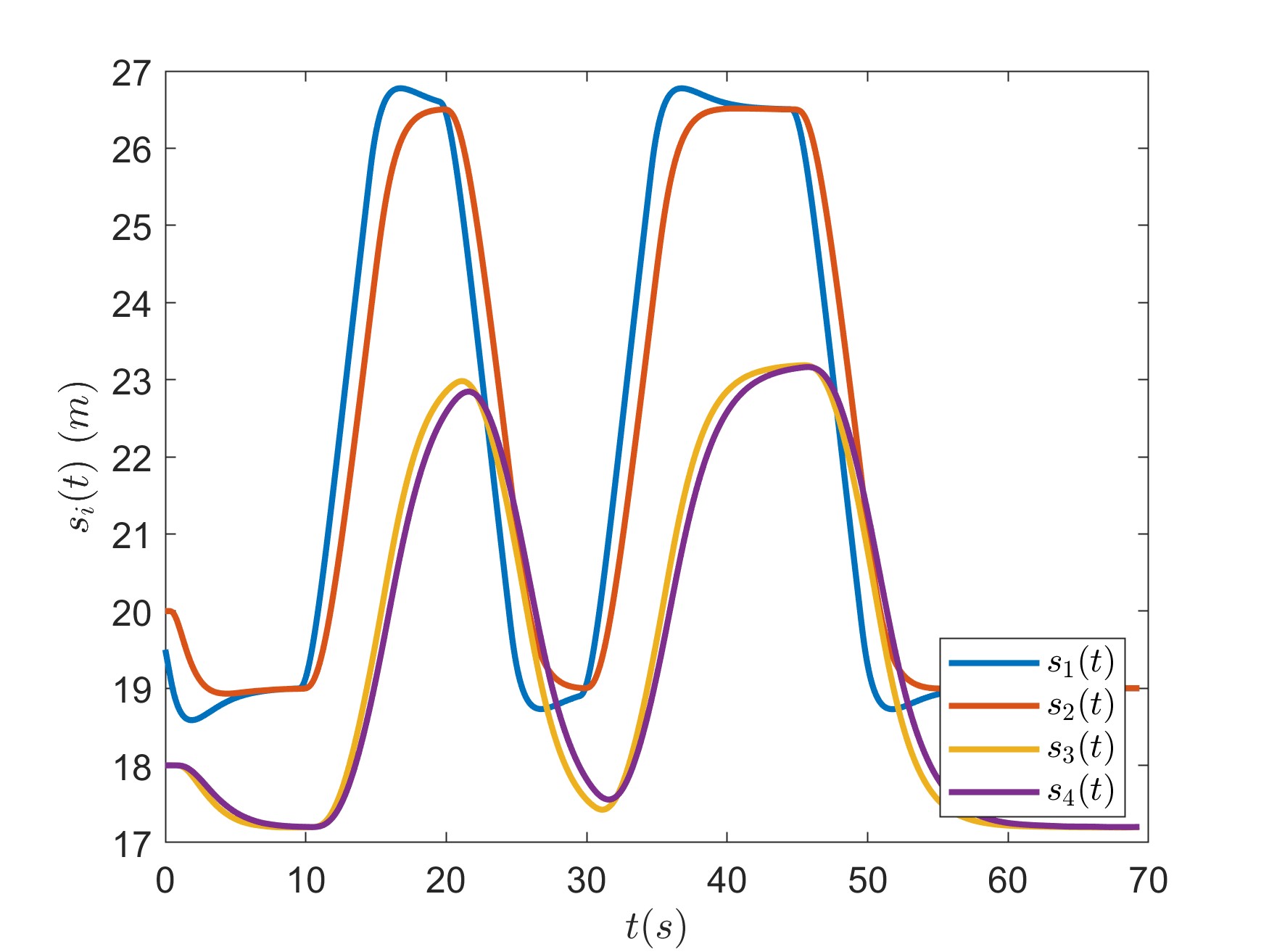}
			\caption{Acceleration (top), speed (middle), and spacing (bottom) of five vehicles, with dynamics described by (\ref{dy1})--(\ref{dy3}), under the predictor-based CACC laws (\ref{CLD}), (\ref{8}), where the selected parameters and control gains are shown in Table \ref{table2} and Table \ref{table3}, respectively. Initial conditions are $v_{i_0} = 10 \left(\frac{m}{s} \right)$, $i = 1,2,3,4$, $v_{{\rm l}_0} = 9 \left(\frac{m}{s} \right) $; $s_{i_0} = h_{i}\times 10 + r_i$~$(m)$, $i = 2, 3, 4$, $s_{1_0} = 9.5+r_1 \nobreakspace (m)$; and $a_i(0)=0$; $u_{i}(s) \equiv 0$, for each vehicle $i$.}
		\end{center}
		\label{Fig2}
	\end{figure}
\section{Experimental Implementation and Validation}
\subsection{Description of the Experimental Setup}
\definecolor{mycolor1}{rgb}{0.00000,0.44700,0.74100}%
\definecolor{mycolor2}{rgb}{0.85000,0.32500,0.09800}%
\newcommand{\lineLeader}{\raisebox{2pt}{\tikz{\draw[color=mycolor1, line width=1.5pt] (0,0) -- (5mm,0);}}}
\newcommand{\lineEgo}{\raisebox{2pt}{\tikz{\draw[color=mycolor2, line width=1.5pt] (0,0) -- (5mm,0);}}}

\begin{figure}
\begin{center}
\includegraphics[width = \linewidth]{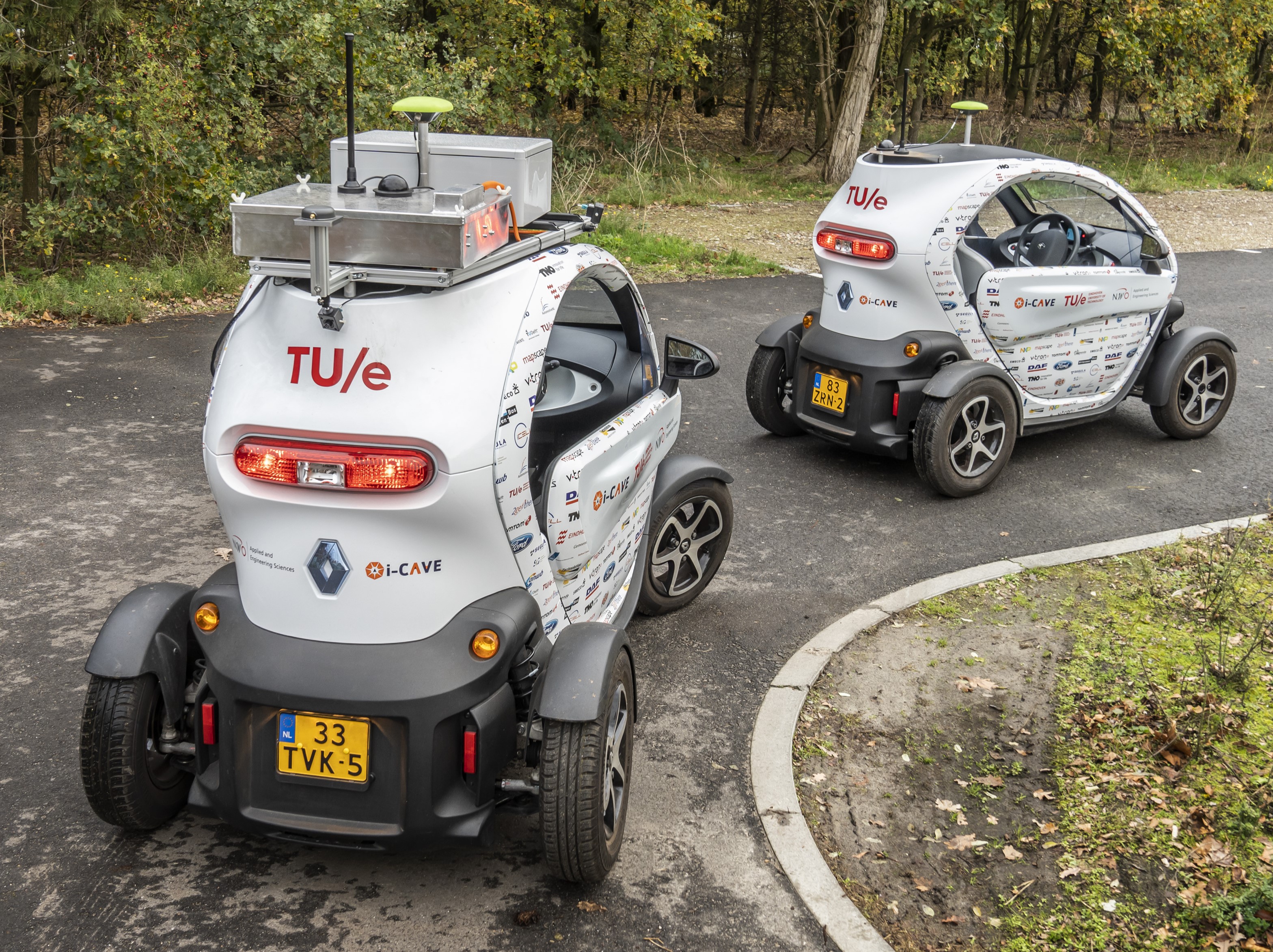}
\caption{Platoon of experimental vehicles that are used to perform the experiments.}
\label{fig:experimental-platoon}
\end{center}
\end{figure}

To experimentally validate the controller, we use a platoon of two full-scale electric vehicles as shown in Fig.~\ref{fig:experimental-platoon}. Details on the automation of the vehicles and associated sensors for a (predictor-based) CACC controller, can be found in \cite{dutch3} and \cite{dutch2}, respectively. The resulting experimental platform exhibits drive-by-wire functionality that results in a longitudinal vehicle response according to \eqref{dy3}, with lag $\tau_i = 0.067 \text{ s}$ and input delay $D_i = 0.15 \text{ s}$. To create distinct input delays for both vehicles, we add an additional input delay in the software of the ego vehicle, to obtain an input delay $D_1 = 0.3 \text{ s}$. The resulting parameters of the experimental setup are identical to vehicles 0 and 1 in Table~\ref{table3}.

To implement the predictor-feedback controller, the measurements $\bar{x}_{ik}$ from \eqref{C4} and $u_{{{i-1},m}_k}$ are required. The inter-vehicle distance $s_i$ is directly measured by the automotive radar. To obtain the ego vehicle's longitudinal velocity $v_i$, the rear-axle rotational speed of the vehicle is measured. The longitudinal acceleration $a_i$ of the vehicle is obtained through an Inertial Measurement Unit (IMU). The exact specifications of the sensors can be found in \cite{dutch}. The preceding vehicle's velocity $v_{i-1}$ and acceleration $a_{i-1}$, as well as its control input $u_{i-1}$ are obtained through V2V communication. During the experiments, the mode of the experienced communication latency $D_{c,{i-1}}$ was 20 ms.

\subsection{Experimental Results}
\begin{figure}
\begin{center}
\begin{subfigure}{0.9\linewidth}
\input{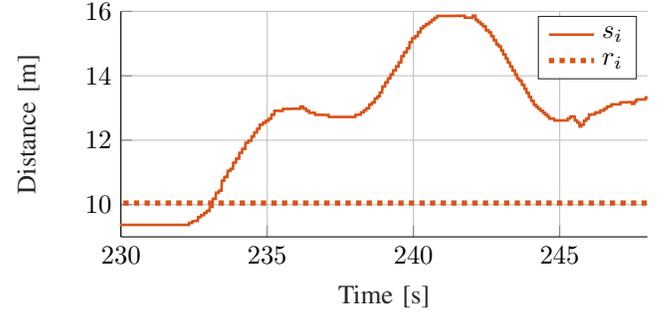}
\caption{Inter vehicle distance $s_i$, as measured by radar.}
\label{subfig:distance-response}
\end{subfigure}
\begin{subfigure}{0.9\linewidth}
\vspace{2mm}
\input{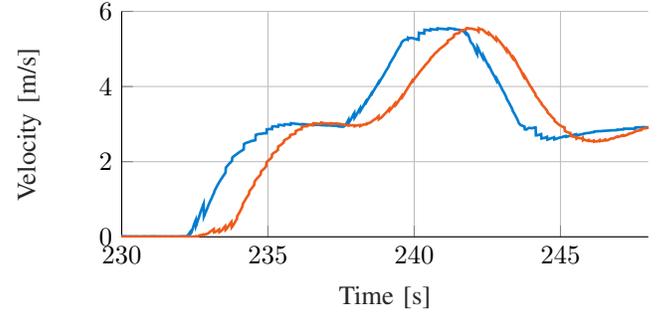}
\caption{Platoon velocity response.}
\end{subfigure}
\begin{subfigure}{0.9\linewidth}
\vspace{2mm}
\input{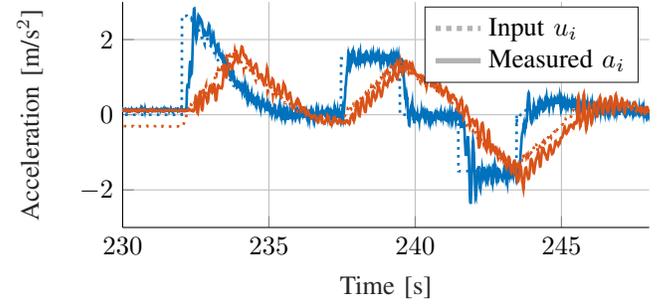}
\caption{Platoon acceleration response.}
\end{subfigure} 
\caption{Measured experimental response of vehicle (\protect\lineEgo) deploying predictor-based controller \eqref{CLD}, \eqref{8}, to follow leader vehicle (\protect\lineLeader). The ego and leader vehicles' parameters and tuning correspond to vehicles 0 and 1 from Table~\ref{table2}, respectively.}
\label{fig:experimental-response}
\end{center}
\end{figure}

To perform the experiments, the vehicles are initialized in an equilibrium position with both vehicles at standstill. Next, the leader vehicle is controlled to a cruise speed of $v_{i-1} = 3 \left( \frac{m}{s}\right) $, while the follower vehicle is controlled with the predictor-based CACC controller \eqref{CLD}, with the tuning of vehicle 1 from Table~\ref{table2}. Once the cruise speed is reached, positive and negative acceleration steps with an amplitude of $1.5 \left( \frac{m}{s^2}\right) $ are prescribed to the leader vehicle. The measured velocity and acceleration response of the platoon is shown in Fig.~\ref{fig:experimental-response}.

At the start of the experiment (from 230 to 232.5 seconds), a negative acceleration setpoint of the ego vehicle can be observed. This negative acceleration setpoint is the result of the initialization of the platoon, which has a slightly smaller inter-vehicle distance than the desired inter-vehicle distance at standstill, as can be seen in Fig.~\ref{subfig:distance-response}. Since the experimental vehicle cannot drive backward (without physically changing the gear), this results in a negative acceleration setpoint which keeps the ego vehicle stationary until the leader vehicle starts driving. 
The measured experimental response indicates the controller and adopted tuning indeed result in string stability, as the there is essentially no overshoot in both the velocity and acceleration response. In steady state situations, when the leader vehicle is driving with constant velocity, the ego vehicle's speed converges to the leader vehicle's speed. Consequently, the experimental results confirm the theoretical results and show that the controller can operate in practice.

\section{Conclusions}
We provided experimental results, validating in actual implementations, the delay-compensating CACC design from \cite{predictor_based}. The implementation relied on explicit, zero-order hold formulae of the predictor-based CACC law, which was employed in the case of a pair of vehicles with different actuation delays. We also provided consistent simulation results, also studying numerically vehicle and string stability, depending on control/model parameters. Both the simulation and experimental results confirm the effectiveness of the design in guaranteeing vehicle stability, string stability, and tracking, consistently with the respective theoretical results from \cite{predictor_based}. 
\section*{Appendix A}
\numberwithin{equation}{section}
\renewcommand{\theequation}{A.\arabic{equation}}
\setcounter{equation}{0}

In order to study string stability of speed errors propagation, we first recall the $\hat{G}_i(s)$ from \cite{predictor_based}, which is given by
\begin{align}
	\hat{G}_i(s)=\frac{\delta_{i}(s)}{s^3+\left(\frac{1}{\tau_i}+c_i \right) s^2+(\alpha_i+b_i)s+\frac{\alpha_i}{h_i}},
	\label{A.24}
\end{align}
where
\begin{align}
	\delta_{i}(s)=&{\rm e}^{-(D_{i}-D_{i-1})s}  \nonumber \\
	&\times\left(\left(\frac{\alpha_i}{h_i} \tau_{i-1}\left(D_i-D_{i-1} \right) +b_i\tau_{i-1}-\frac{\alpha_i}{h_i}\tau_{i-1}^2 \right.\right.\nonumber\\
	&\left.\left.+ {\rm e}^{-\frac{\left( D_{i}-D_{i-1}\right)}{\tau_{i-1}}}\left(\frac{\alpha_i}{h_i}\tau_{i-1}^2-b_i\tau_{i-1}+c_i \right)  \right)s^2 \right.\nonumber\\
	&\left.+ \left( \left(D_i - D_{i-1} \right) \frac{\alpha_i}{h_i} + b_i\right) s+\frac{\alpha_i}{h_i}\right).
\end{align}
Then by applying Tustin approximation \cite{tustin} and substituting $s=\frac{2}{T_s}\frac{z-1}{z+1}$, we derive its discrete-time counterpart
\begin{align}
	G_i(z)=z^{-(l_i-l_{i-1})}\frac{f_{1,i}z^3+f_{2,i}z^2+f_{3,i}z+f_{4,i}}{g_{1,i}z^3+g_{2,i}z^2+g_{3,i}z+g_{4,i}},
	\label{Gz}
\end{align}
where
\begin{align}
	f_{1,i}=&4 \left(\frac{\alpha_i}{h_i} \tau_{i-1}\left(D_i-D_{i-1} \right) +b_i\tau_{i-1}-\frac{\alpha_i}{h_i}\tau_{i-1}^2 \right.\nonumber\\
	&\left.+ {\rm e}^{-\frac{\left( D_{i}-D_{i-1}\right)}{\tau_{i-1}}}\left(\frac{\alpha_i}{h_i}\tau_{i-1}^2-b_i\tau_{i-1}+c_i \right)  \right) h_i \tau_i T_s \nonumber\\
	&+ 2 \left( \left(D_i - D_{i-1} \right) \frac{\alpha_i}{h_i} + b_i\right) h_i \tau_i T_s^2 + \alpha_i \tau_i T_s^3,\\
	f_{2,i}=&-4 \left(\frac{\alpha_i}{h_i} \tau_{i-1}\left(D_i-D_{i-1} \right) +b_i\tau_{i-1}-\frac{\alpha_i}{h_i}\tau_{i-1}^2 \right.\nonumber\\
	&\left.+ {\rm e}^{-\frac{\left( D_{i}-D_{i-1}\right)}{\tau_{i-1}}}\left(\frac{\alpha_i}{h_i}\tau_{i-1}^2-b_i\tau_{i-1}+c_i \right)  \right) h_i \tau_i T_s \nonumber\\
	&+ 2 \left( \left(D_i - D_{i-1} \right) \frac{\alpha_i}{h_i} + b_i\right) h_i \tau_i T_s^2 + 3 \alpha_i \tau_i T_s^3,\\
	f_{3,i}=&-4 \left(\frac{\alpha_i}{h_i} \tau_{i-1}\left(D_i-D_{i-1} \right) +b_i\tau_{i-1}-\frac{\alpha_i}{h_i}\tau_{i-1}^2 \right.\nonumber\\
	&\left.+ {\rm e}^{-\frac{\left( D_{i}-D_{i-1}\right)}{\tau_{i-1}}}\left(\frac{\alpha_i}{h_i}\tau_{i-1}^2-b_i\tau_{i-1}+c_i \right)  \right) h_i \tau_i T_s\nonumber\\
	& - 2 \left( \left(D_i - D_{i-1} \right) \frac{\alpha_i}{h_i} + b_i\right) h_i \tau_i T_s^2 + 3 \alpha_i \tau_i T_s^3,\\
	f_{4,i}=&4 \left(\frac{\alpha_i}{h_i} \tau_{i-1}\left(D_i-D_{i-1} \right) +b_i\tau_{i-1}-\frac{\alpha_i}{h_i}\tau_{i-1}^2 \right.\nonumber\\
	&\left.+ {\rm e}^{-\frac{\left( D_{i}-D_{i-1}\right)}{\tau_{i-1}}}\left(\frac{\alpha_i}{h_i}\tau_{i-1}^2-b_i\tau_{i-1}+c_i \right)  \right) h_i \tau_i T_s\nonumber\\
	& - 2 \left( \left(D_i - D_{i-1} \right) \frac{\alpha_i}{h_i} + b_i\right) h_i \tau_i T_s^2 + \alpha_i \tau_i T_s^3,\\
	g_{1,i}=&8 h_i \tau_i + 4 h_i T_s + 4 c_i h_i \tau_i T_s + 2 \alpha_i h_i \tau_i T_s^2 \nonumber\\
	&+ 2 b_i h_i \tau_i T_s^2 + \alpha_i \tau_i T_s^3,\\
	g_{2,i}=&-24 h_i \tau_i - 4 h_i T_s - 4 c_i h_i \tau_i T_s + 2 \alpha_i h_i \tau_i T_s^2 \nonumber\\
	&+ 2 b_i h_i \tau_i T_s^2 + 3 \alpha_i \tau_i T_s^3,\\
	g_{3,i}=&24 h_i \tau_i - 4 h_i T_s - 4 c_i h_i \tau_i T_s - 2 \alpha_i h_i \tau_i T_s^2 \nonumber\\
	&- 2 b_i h_i \tau_i T_s^2 + 3 \alpha_i \tau_i T_s^3,\\
	g_{4,i}=&-8 h_i \tau_i + 4 h_i T_s + 4 c_i h_i \tau_i T_s - 2 \alpha_i h_i \tau_i T_s^2 \nonumber\\
	&- 2 b_i h_i \tau_i T_s^2 + \alpha_i \tau_i T_s^3.
\end{align}


\begin{thebibliography}{00}
	\bibitem{nominal_control_1}  
	E. Abolfazli, B. Besselink, and T. Charalambous, “Minimum time headway in platooning systems under the MPF topology for different wireless communication scenarios,” {\em IEEE Transactions on Intelligent Transportation Systems}, vol. 24, pp. 4377–4390, 2023.
	
	
	\bibitem{predictor_feedback1}  
	N. Bekiaris-Liberis, “Robust string stability and safety of CTH predictor-feedback CACC,” {\em IEEE Transactions on Intelligent Transportation Systems}, vol. 24, pp. 8209–8221, 2023.  
	
	
	\bibitem{predictor_feedback2}  
	N. Bekiaris-Liberis, C. Roncoli, and M. Papageorgiou, “Predictor-based adaptive cruise control design,” {\em IEEE Transactions on Intelligent Transportation Systems}, vol. 19, pp. 3181–3195, 2018.  
	
	\bibitem{predictor-based_implementation2}
	A. Bertino, P. Naseradinmousavi and M. Krstić, “Delay-adaptive control of a 7-DOF robot manipulator: design and experiments,” {\em IEEE Transactions on Control Systems Technology}, vol. 30, no. 6, pp. 2506--2521, 2022. 
	
	\bibitem{nominal_control_2}  
	Y. Bian, Y. Zheng, W. Ren, S. Eben Li, J. Wang, and K. Li, “Reducing time headway for platooning of connected vehicles via V2V communication,” {\em Transportation Research Part C: Emerging Technologies}, vol. 102, pp. 87–105, 2019.  
	
	\bibitem{speed_error1}  
	A. Bose and P. A. Ioannou, “Analysis of traffic flow with mixed manual and semiautomated vehicles,” {\em IEEE Transactions on Intelligent Transportation Systems}, vol. 4, pp. 173–188, 2003.  
	
	\bibitem{b31}  
	B. Caiazzo, D. G. Lui, A. Mungiello, A. Petrillo, and S. Santini, “On the resilience of autonomous connected vehicle platoons under DoS attacks: A predictor-based sampled data control,” in {\em IEEE Int. Conf. Intell. Transp. Syst.}, Bilbao, Spain, pp. 4907–4912, 2023.  
	\bibitem{b6}  
	L. C. Davis, “Method of compensation for the mechanical response of connected adaptive cruise control vehicles,” {\em Physica A: Statistical Mechanics and its Applications}, vol. 562, p. 125402, 2021.  
	
	\bibitem{speed_error2}  
	J. I. Ge and G. Orosz, “Dynamics of connected vehicle systems with delayed acceleration feedback,” {\em Transportation Research Part C: Emerging Technologies}, vol. 46, pp. 46–64, 2014.  
	
	\bibitem{dutch}  
	R. de Haan, T. P. J. van der Sande, E. Lefeber, and I. J. M. Besselink, “Cooperative adaptive cruise control for heterogeneous platoons with delays: Controller design and experiments,” {\em IEEE Transactions on Control Systems Technology}, in press, 2024.  
	
	\bibitem{dutch1}  
	R. de Haan, T. van der Sande, and E. Lefeber, “Observer-based cooperative adaptive cruise control for heterogeneous vehicle platoons with actuator delay,” in {\em IEEE International Conference on Intelligent Transportation Systems}, Bilbao, Spain, pp. 5204–5209, 2023.  
	
	\bibitem{dutch2}  
	R. de Haan, L. Redi, T. van der Sande, and E. Lefeber, “Platooning of heterogeneous vehicles with actuation delays: Experimental results,” {\em IFAC-PapersOnLine}, vol. 58, no. 27, pp. 131–136, 2024.  
	
	\bibitem{dutch3}
	F. N. Hoogeboom, {\em Safety of Automated Vehicles:} {\em Design, Implementation, and Analysis}. Ph.D. thesis, Eindhoven University of Technology, 2020.
	
	\bibitem{b7}  
	S. Huang and W. Ren, “Autonomous intelligent cruise control with actuator delays,” {\em J. of Intell. \& Robot. Syst.}, vol. 23, pp. 27–43, 1998.  
	
	\bibitem{tustin}
	K. B. Janiszowski, “A modification and the Tustin approximation,” {\em IEEE Transactions on Automatic Control}, vol. 38, no. 8, pp. 1313--1316, 1993.
	
	\bibitem{predictor-based_implementation1}
	M. Jankovic and S. Magner, “Disturbance attenuation in time-delay systems — A case study on engine air-fuel ratio control,” {\em Proceedings of the 2011 American Control Conference}, San Francisco, CA, USA, pp. 3326--3331, 2011.
	
	\bibitem{Discrete_predictor}
	I. Karafyllis and M. Krstic, “Nonlinear stabilization under sampled and delayed measurements, and with inputs subject to delay and zero-order hold”, {\em IEEE Transactions on Automatic Control}, vol. 57, no. 5, pp. 1141--1154, 2012.
	
	
	\bibitem{b14}  
	T. G. Molnar, W. B. Qin, T. Insperger, and G. Orosz, “Application of predictor feedback to compensate time delays in connected cruise control,” {\em IEEE Trans. Intell. Transp. Syst.}, vol. 19, pp. 545–559, 2018.  
	
	\bibitem{prediction1}  
	T. G. Molnar, A. K. Kiss, A. D. Ames, and G. Orosz, “Safety-critical control with input delay in dynamic environment,” {\em IEEE Transactions on Control Systems Technology}, vol. 31, pp. 1507–1520, 2023. 
	
	
	\bibitem{b17}  
	S. Öncü, J. Ploeg, N. van de Wouw, and H. Nijmeijer, “Cooperative adaptive cruise control: Network-aware analysis of string stability,” {\em IEEE Trans. on Intell. Transp. Syst.}, vol. 15, pp. 1527–1537, 2014.  
	
	
	\bibitem{italian}  
	A. Petrillo, A. Salvi, S. Santini, and A. S. Valente, “Adaptive multi-agent synchronization for collaborative driving of autonomous vehicles with multiple communication delays,” {\em Transportation Research Part C: Emerging Technologies}, vol. 86, pp. 372–392, 2018.  
	
	\bibitem{L_stability}  
	J. Ploeg, N. van de Wouw, and H. Nijmeijer, “Lp string stability of cascaded systems: Application to vehicle platooning,” {\em IEEE Transactions on Control Systems Technology}, vol. 22, pp. 786–793, 2014.  
	
	\bibitem{predictor_feedback3}  
	A. Samii and N. Bekiaris-Liberis, “Simultaneous compensation of actuation and communication delays for heterogeneous platoons via predictor-feedback CACC with integral action,” {\em IEEE Transactions on Intelligent Vehicles}, in press, 2024.  
	
	\bibitem{predictor_based}  
	A. Samii and N. Bekiaris-Liberis, “Predictor-based CACC design for heterogeneous vehicles with distinct input delays,” {\em IEEE Open Journal of Intelligent Transportation Systems}, vol. 5, pp. 783–796, 2024.  
	
	\bibitem{predictor-based_implementation3}
	S. M. Schlanbusch, J. Zhou and R. Schlanbusch, “Quantized predictor-based tracking control for input delay systems: application to a helicopter system,” {\em IEEE 63rd Conference on Decision and Control (CDC)}, Milan, Italy, pp. 2226--2231, 2024.
	
	
	\bibitem{b13}  
	M. Wang et al., “Delay-compensating strategy to enhance string stability of autonomous vehicle platoons,” {\em Transportmetrica B: Transport Dynamics}, vol. 6, pp. 211–229, 2016.  
	
	
	\bibitem{b29} 
	L. Xiao and F. Gao, “Practical string stability of platoon of adaptive cruise control vehicles,” {\em IEEE Transactions on intelligent transportation systems}, vol. 12, pp. 1184–1194, 2011.
	
	\bibitem{dy1} 
	H. Xing, J. Ploeg, and H. Nijmeijer, “Smith predictor compensating for vehicle actuator delays in cooperative ACC systems,” {\em IEEE Transactions on Vehicular Technology}, vol. 68, pp. 1106–1115, 2018.
	
	
	\bibitem{dy2} 
	H. Xing, J. Ploeg, and H. Nijmeijer, “Compensation of communication delays in a cooperative ACC system,” {\em IEEE Transactions on Vehicular Technology}, vol. 69, pp. 1177–1189, 2019.
	
	\bibitem{b38} 
	D. Yanakiev and I. Kanellakopoulos, “Longitudinal control of automated CHVs with significant actuator delays,” {\em IEEE Transactions on Vehicular Technology}, vol. 50, pp. 1289–1297, 2001.
	
	\bibitem{b50} 
	C. Zhao and H. Yu, “Robust safety for mixed-autonomy traffic with delays and disturbances,” {\em IEEE Transactions on Intelligent Transportation Systems}, in press, 2024.
	
	\bibitem{speed_error3} 
	Y. Zhang, Y. Bai, J. Hu, D. Cao, and M. Wang, “Memory-anticipation strategy to compensate for communication and actuation delays for string-stable platooning,” {\em IEEE Transactions on Intelligent Vehicles}, vol. 8, pp. 1145–1155, 2022.
	
	\bibitem{distinc_input_delay2} 
	H. Zhang, J. Liu, Z. Wang, C. Huang, and H. Yan, “Adaptive switched control for connected vehicle platoon with unknown input delays,” {\em IEEE Transactions on Cybernetics}, vol. 53, pp. 1511–1521, 2023.
	
\end{thebibliography}
\end{document}